\documentclass{article}
\usepackage[left=1in, right=1in, top=1in, bottom=1in]{geometry}
\usepackage{cite,graphicx,subfigure,setspace}
\usepackage[small,bf,up]{caption}

\begin{document}
\title{Nanobeam photonic crystal cavity quantum dot laser}
\author{Yiyang Gong*, Bryan Ellis, Gary Shambat, Tomas Sarmiento,\\ James S. Harris, and Jelena Vu\v{c}kovi\'{c} \\
    \small\textit{Department of Electrical Engineering, Stanford University, Stanford, CA 94305} \\
    \small\textit{*email:yiyangg@stanford.edu}}
%\twocolumn[
%\begin{@twocolumnfalse}
\maketitle
\begin{abstract}
The lasing behavior of one dimensional GaAs nanobeam cavities with embedded InAs quantum dots is studied at room temperature. Lasing is observed throughout the quantum dot PL spectrum, and the wavelength dependence of the threshold is calculated. We study the cavity lasers under both 780 nm and 980 nm pump, finding thresholds as low as 0.3 $\mu$W and 19 $\mu$W for the two pump wavelengths, respectively. Finally, the nanobeam cavity laser wavelengths are tuned by up to 7 nm by employing a fiber taper in near proximity to the cavities. The fiber taper is used both to efficiently pump the cavity and collect the cavity emission.
\end{abstract}
%\ocis{(140.3410) Laser resonators, (230.5298) Photonic crystals, (230.5590) Quantum-well, -wire and -dot devices, (230.2285) Fiber devices and optical amplifiers, (140.3600) Lasers, tunable}
%\end{@twocolumnfalse}
%]
%\doublespacing

Photonic crystal (PC) cavities have had a large impact in the field of low threshold lasers \cite{Painter_PClaser,Marko_PClaser}, as these high quality ($Q$-) factor and small mode volumes ($V_{m}$) resonators reduce the lasing threshold by reducing the amount of active material in the cavity and enhancing the spontaneous emission rate of the emitters through Purcell enhancement, which is proportional to $Q/V_{m}$ \cite{Purcell}. In addition, the high Purcell factors and small mode volumes (i.e., large photon densities) of PC cavities also enable lasers with very fast direct modulation speeds exceeding 100 GHz \cite{Hatice_arraylaser,Dirk_laserrateeq}, which have the potential to be used in opto-electronic communications. Finally, the use of quantum dot active material inside PC cavity lasers to further lower the lasing threshold (by minimizing non-radiative surface recomination effects) has been studied \cite{Bryan_laser,Arakawa_RTlasing,Strauf_QDlaser,Arakawa_SingleQDlaser}. Other potential applications of these devices include compact chemical or mass sensors \cite{Loncar_lasersensing}.

In this work, we study the lasing properties of GaAs nanobeam cavities with InAs quantum dots (QDs) at room temperature. One dimensional (1D) nanobeam cavities have been previously designed and fabricated in a variety of passive materials, such as Si \cite{MIT_1D,Marko_Si1D}, Si$_{3}$N$_{4}$ \cite{Painter_1Dmodes} and SiO$_{2}$ \cite{Gong_quartz1D}. Very recently, lasing in such cavities incorporating multi-quantum well material has been demonstrated \cite{Loncar_beamlaser}. Because of their small footprint, such 1D PC cavities also have potential as compact light sources for on-chip optical communications, and proposals for employing nanomechanical properties of such structures to build tunable lasers have been made \cite{Painter_tuneablelaser}. Much like two dimensional PC cavities, nanobeam cavities have high $Q$ and low $V_{m}$, thereby greatly decreasing the lasing threshold via the Purcell enhancement of spontaneous emission rate. The nanobeam laser are also suitable for coupling to fiber taper, which is demonstrated in this work. We employ this coupling to simultaneously tune of the lasing wavelength by up to 7 nm by controlling the overlap between the cavity and the taper.

\begin{figure}[hbtp]
\centering
\includegraphics[width=2.8in]{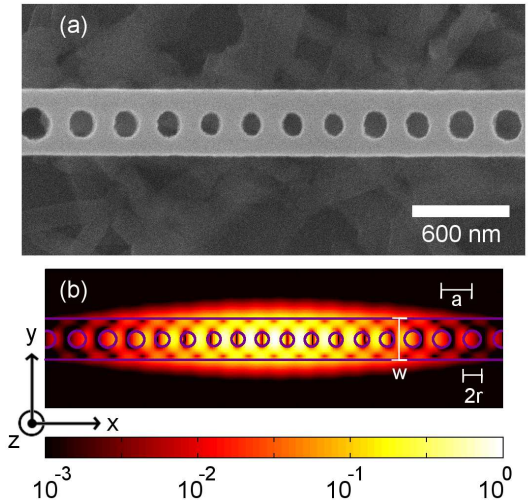}
\caption{(a) The fabricated 1D nanobeam cavity. (b) The electric field intensity ($|E|^2$) of the fundamental mode supported by the cavity.}
\label{fig:modes}
\end{figure}

The cavity design is based on a beam having thickness $d$ and width $w$, and circular holes are patterned along the beam with period $a$ and radii $r=0.3a$ in the photonic crystal mirror region (Fig. \ref{fig:modes}(a)). The cavity comprises of holes spaced at $a'=0.84a$ at the center of the cavity, and holes size of $r'=0.84r$. The hole spacing and size increase parabolically from the center of the cavity outwards, extending 6 holes on either side of the cavity. The cavity is designed with $d=0.7a$ and $w=1.3a$, and is simulated by the three dimensional finite difference time domain (3D-FDTD) method with 20 units per lattice constant and perfectly matched layer (PML) absorbing boundary conditions. We compute the $Q$ of the cavity using $Q=\omega U/P$, where $\omega$ is the frequency of the cavity, $U$ is the total energy of the mode, and $P$ is the time-averaged energy radiated transverse to the beam length (i.e. not through the ends of the beams, where the leakage is suppressed by the distributed Bragg reflection). Using the FDTD simulation, we find the $|E|^2$ field profile of the fundamental transverse electric (TE-) like cavity mode shown in Fig \ref{fig:modes}(b), which is dominated by the E$_{y}$ component. We also find that further increase in the number of photonic crystal mirror layers beyond 15 did not increase the overall $Q$ of the cavity. Finally, from the simulations, we obtain $Q=3.5\times 10^{4}$, $V_{m}=0.8 (\lambda/n)^3$, and normalized frequency of $a/\lambda=0.25$. We work with the fundamental TE mode as it has the highest $Q$ and lowest $V_{m}$ among all TE modes. In experiment, we observe a significant reduction in $Q$ relative to theoretical prediction (by a factor of 3), resulting from fabrication imperfections (such as edge roughness from the dry etch and lithographic tolerances to the hole position), or from absorption losses in the QDs and the wetting layer.

The employed membrane structure was grown by molecular beam epitaxy (MBE) and consists of a 1 $\mu$m Al$_{0.8}$Ga$_{0.2}$As sacrificial layer and a 240 nm GaAs membrane that contains three layers of InAs quantum dots separated by 50 nm GaAs spacers. To achieve emission at 1.3 $\mu$m, the dots were capped with a 6 nm In$_{0.15}$Ga$_{0.85}$As strain-reducing layer. The quantum dots were formed by depositing 2.8 monolayer (ML) of InAs at 510$^{\circ}$C using a growth rate of 0.05 ML/s. These growth conditions result in a dot density of $3\times 10^{10}$ cm$^{-2}$, as estimated from atomic force microscopy (AFM) measurements of uncapped quantum dot samples. Fabrication of the PC cavities is done by spinning a 300 nm layer of the electron beam resist ZEP-520A on top of the wafer, and performing e-beam lithography. The written pattern is then transferred to the GaAs membrane using a Cl$_{2}$:BCl$_{3}$ dry etch. Finally, the nanobeam is undercut with a 7\% HF solution in water. An example fabricated structure is shown in Fig. \ref{fig:modes}(a).

\begin{figure}[hbtp]
\centering
\includegraphics[width=3.3in]{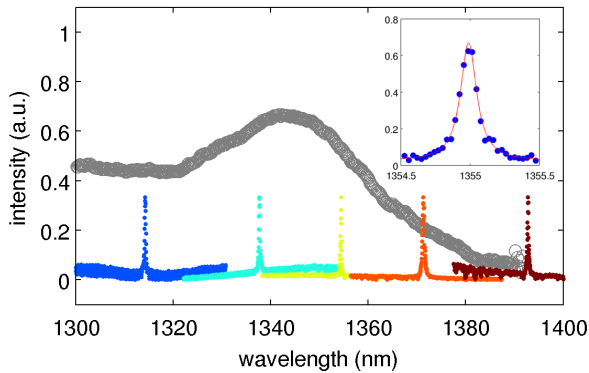}
\caption{Normalized PL spectra from representative cavities above lasing threshold (colored points). The PL spectrum from QDs in bulk (unpatterned film) is also shown (gray circles). The inset shows a zoomed-in cavity spectrum below lasing threshold and its fit to a Lorentzian lineshape, corresponding to $Q=9,700$.}
\label{fig:cavs}
\end{figure}

We pump the cavities at both 780 nm (above the GaAs band gap positioned at $\lambda_{g}=870$nm) and at 980nm (below $\lambda_{g}$, but above the emission wavelength of the QDs), both at room temperature. Suspended bridge nanobeam cavities have been shown to have very small heat conduction \cite{Notomi_optbi}. However, the 980 nm pumping avoids heating of the cavity at high pump powers, and allows high power continuous wave (CW) pumping. The CW pump laser is focused onto the beam from normal incidence with a 100$\times$ objective lens with numerical aperture NA=0.5. The photoluminescence (PL) from the sample is also collected from the direction perpendicular to the plane of the chip and sent to a spectrometer with an InGaAs CCD array. The PL from QDs in a unpatterned region of the sample is shown in Fig. \ref{fig:cavs}. The PL spectra of various cavities with slightly different lattice constants and radii are also shown in Fig. \ref{fig:cavs}, with pump powers above threshold with the 980 nm pump. The Lorentzian fit to a cavity spectrum below threshold with $Q = 9,700$ is shown in the inset of Fig. \ref{fig:cavs}. 

%2010/01/07 cavs.m
%inset 2010/01/22 runs1/13.9  divide by run3/bulk
%bulk from 2009/12/26 run7/8.66E-5

\begin{figure}[hbtp]
\centering
\includegraphics[width=3.3in]{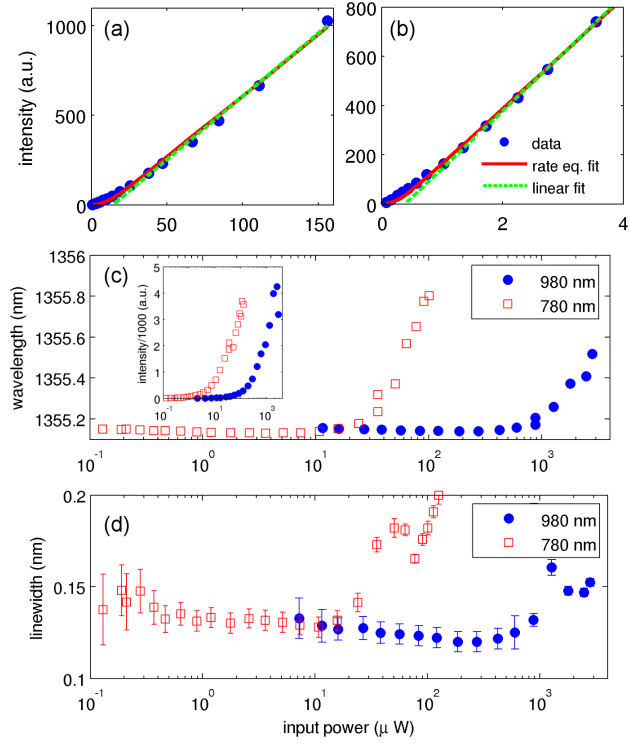}
\caption{The light-in light-out curves of a representative cavity, using (a) the 980 nm pump and (b) the 780 nm pump laser. Fits from the rate equations, and linear fits to the above threshold behavior are also shown. (c) The power dependence of the cavity wavelength with 780 nm and 980 nm pump. The red-shift at high pump powers indicates structure heating, and it kicks off sooner if above-GaAs bandgap laser (780nm) is employed, as expected. The inset shows the cavity intensity for larger pump powers, where the beginning of saturation is observed toward the end of both traces. (d) The power dependence of the cavity linewidth with 780 nm and 980 nm pump. The pump power (horizontal axis) is measured before the objective in all cases.}
\label{fig:power}
\end{figure}

We also study the pump power dependence of cavities by varying the pump power of an unchopped CW 980 nm laser from as little as 0.1 $\mu$W to as much as 10mW. The output power of the nanobeam laser as a function of the pump power is shown in Fig. \ref{fig:power}(a). The experiment is repeated with the same cavity, but with the 780 nm pump (Fig. \ref{fig:power}(b)). We fit the data to the standard rate equations \cite{ColdrenCorzine}:
\begin{eqnarray}
\frac{dN}{dt}&=&\eta \frac{L_{in}}{\hbar \omega V_{a}}-N\left(\frac{F}{\tau_{r}}+\frac{1}{\tau_{nr}}\right)-v_{g}gP \\
\frac{dP}{dt}&=&\Gamma v_{g} g P + \Gamma \beta \frac{N}{\tau_{r}}-\frac{P}{\tau_{p}} ,
\end{eqnarray}
where $N(P)$ represents the carrier (photon) density, $F$ is the Purcell factor, $\eta$ represents the fraction of incident pump power ($L_{in}$) absorbed in the active region, $V_{a}$ is the active volume of the laser, $\tau_{r}$ ($\tau_{nr}$) is the radiative (non-radiative) recombination lifetime, $v_{g}=1\times 10^{10}$ cm/s is the group velocity of light in the active medium, $\tau_{p}$ is the photon lifetime of the cavity, $\beta$ is the fraction of spontaneous emission coupled to the cavity mode, and $\Gamma$ is the mode overlap with the QDs.  Because the homogeneous linewidth of such QDs at room temperature is approximately 10 meV \cite{Arakawa_RTlasing}, which far exceeds the cavity linewidth, the Purcell enhancement is negligible ($F\approx 1$) \cite{Woerdman_gamma}. A logarithmic gain model $g=g_{0}$ln$(N/N_{tr})$ is used where $g_{0}$ is the gain coefficient in units of cm$^{-1}$ and N$_{tr}$ is the transparency carrier density in units of cm$^{-3}$.  The photon lifetime $\tau_{p}=Q/\omega$ is estimated from the linewidth of the cavity resonance around threshold to be 7.1 ps.  The QD radiative lifetime in bulk, $\tau_{r}$, is estimated from the literature to be about 3 ns, and the non-radiative lifetime, $\tau_{nr}$, is too long to significantly affect the fits.  We also expect that the non-radiative recombination occurring at the surfaces in our structures is significantly lower than in quantum well lasers, as a result of the spatial confinement of the QD exciton.  This small non-radiative recombination rate, in addition to low threshold, causes a soft turn-on of the laser structures shown in Fig \ref{fig:power}(a).

Since it is difficult to estimate the gain parameters and the fraction of absorbed pump power in our structures, we fit the rate equations with $\beta$, $g_{0}$, N$_{tr}$, and $\eta$ as variable parameters. We simultaneously fit the data from the 780 nm pump and the 980 nm pump to the model with the same  $\beta$, $g_{0}$, and N$_{tr}$, but different $\eta$. The best fit to the data is obtained with $g_{0}=6.2\times 10^{4}$cm$^{2}$ and $N_{tr}=7.9\times 10^{15}$cm$^{-3}$, comparable to previous studies with similar quantum dots \cite{Arakawa_RTlasing}.  For our lasers we find $\beta=0.88$, $\eta=1.3\times 10^{-5}$ for the 980 nm pump, and $\eta=6.3\times 10^{-4}$ for the 780 nm pump. The difference in $\eta$ for the two pump powers is expected, since the 980 nm pump laser has lower energy than the GaAs band gap and therefore is weakly absorbed (only by QDs and the wetting layer).  Despite low Purcell enhancement, a high $\beta$ factor is achieved, resulting from redirection of spontaneous emission into a single mode, similar to vertical nanowire antennas \cite{Gerard_nanowire,Loncar_nanowire}. To find the threshold of the laser we use a linear fit to the light-in light-out curve above threshold, and find the thresholds to be 19 $\mu$W and 0.3 $\mu$W, for the 980 nm and 780 nm pump, respectively. Again, the reduction in threshold highlights the improved pump efficiency with the 780 nm pump. The threshold with the 780 nm is an order of magnitude lower than that reported in Ref. \cite{Arakawa_RTlasing}, where a chopped CW pump was used to reduce heating effects. We observe a reduction in threshold despite not using any chopping, which could come from the reduced number of QDs layers in our structure, or from the high $\beta$-factor of this cavity design.

In addition, we notice that two pump wavelengths create different behavior in the cavity heating. For both pump wavelengths, the cavity wavelength is unchanged at low pump powers, but begins to red shift at high pump powers (Fig. \ref{fig:power}(c)). Moreover, the wavelength shift for the 780 nm pump begins at lower powers than the one for the 980nm pump, which is expected, as the 780nm pump is more efficiently absorbed in the material. We also study the cavity linewidth as a function of pump power, but observe only a small narrowing (Fig. \ref{fig:power}(d)), as the cavity is near the resolution of our spectrometer and small linewidth narrowing at threshold is a signature of high $\beta$ factor lasers \cite{Koch_laser}. There is a noticeable increase in linewidth above 10 $\mu$W pump power associated with heating losses, and again occurring sooner with the 780nm pump. Finally, in the inset of Fig. \ref{fig:power}(c), we show the high pump power dependence of the cavity output intensity, and the end of each trace represents the pump power where the cavity output starts to decrease. While the two pump wavelengths show approximately the same power output, the cavity linewidth is irreversibly broadened with the 780 nm before the saturation behavior (as in Fig. \ref{fig:power}(d)), suggesting heating damage to the cavity. On the other hand, damage to the cavity is not observed with the 980 nm pump.

\begin{figure}[hbtp]
\centering
\includegraphics[width=3.3in]{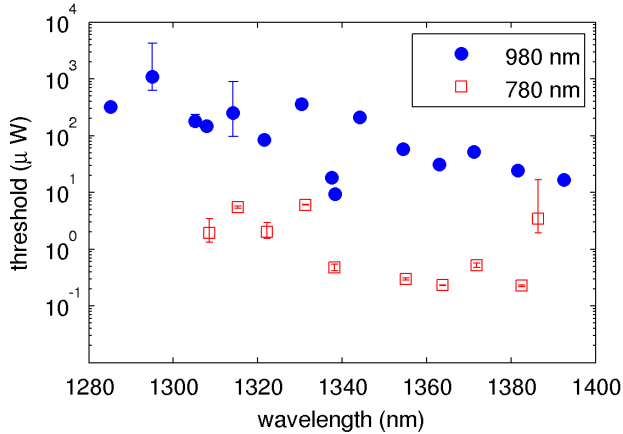}
\caption{The thresholds of various nanobeam lasers obtained by linear fit to the above threshold behavior, using both the 780 nm and the 980 nm pump. Threshold pump powers are measured before the objective lens in all cases.}
\label{fig:thresholds}
\end{figure}

We also investigate multiple cavities throughout the PL spectrum of the QDs, finding each threshold by a linear fit to the above threshold behavior and plotting the results in Fig. \ref{fig:thresholds}. First, we notice that the thresholds increase by nearly an order of magnitude as we move toward the blue side of the PL spectrum, for both the 780 nm and the 980 nm pump. This results from the degradation of the cavity $Q$-factor with decreasing wavelength, as more quantum dots are able to absorb the emission from the cavity. In addition, we also observe that the use of the 780 nm pump always results in lower thresholds than the 980 nm, by approximately two orders of magnitude. Again, this corroborates the fact that the pumping above the GaAs band gap efficiently delivers carriers to the QDs.

\begin{figure}[hbtp]
\centering
\includegraphics[width=3.3in]{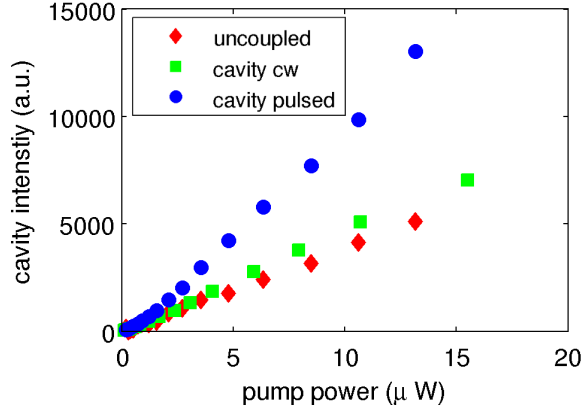}
\caption{The light-in light-out curve for the same cavity as in Fig \ref{fig:power}(a)-(b), pumped with a pulsed 830 nm laser, and by a CW 830 nm laser. The emission from a portion of the PL spectrum not coupled to the cavity is also shown.}
\label{fig:cavpumps}
\end{figure}

In order to check the lossy mechanisms due to heating, we also pump the cavity with a pulsed 830 nm laser (35 ns pulse, 150 ns repetition period). It should be noted that this corresponds to a quasi-CW regime, as the pulse duration is much longer than any recombination time scales of the system, but the modulation helps reduce heating losses. The cavity emission as a function of the peak CW power is plotted in Fig. \ref{fig:cavpumps} for the various cases of CW and pulsed pumping. Lasing is observed in both cases, but the pulsed pump generated a higher slope of the LL curve. Moreover, the saturation at higher pump powers is delayed in the case of the pulsed pumping. This is attributed to the reduction of the heating effect, which leads to higher laser efficiency.

\begin{figure}[hbtp]
\centering
\includegraphics[width=5.5in]{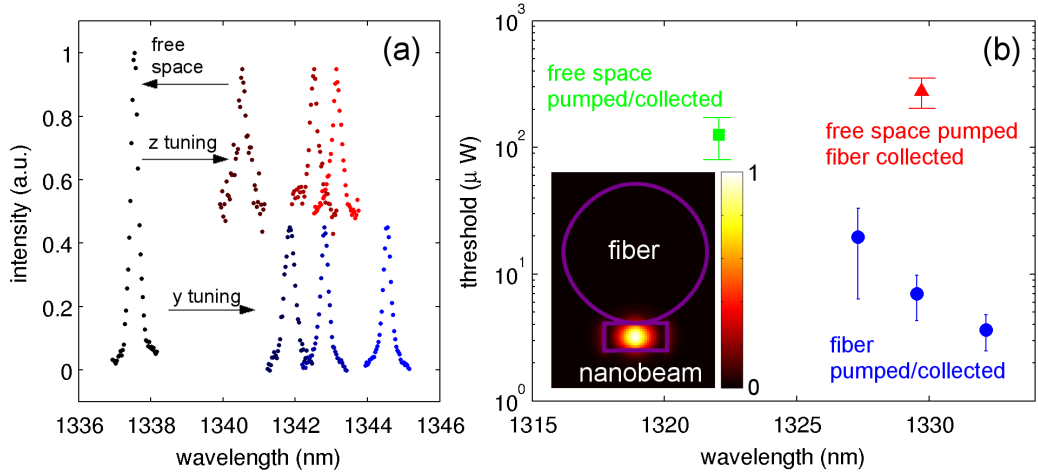}
\caption{(a) Spectra from a nanobeam cavity as it is tuned by the movement of a fiber taper in close proximity to the cavity. The free space spectrum without the fiber taper is shown as a reference, and taper movement in the $y$- and $z$-directions (shown in Fig. \ref{fig:modes}) tunes the cavity mode by over 7 nm. The spectra for the tuned cavity are scaled for clarity. (b) The lasing thresholds of one cavity pumped from free space (normal incidence) and through the fiber taper, with collection through the fiber taper in both cases. The fiber taper position is varied to tune the lasing wavelengths. A reference case without any fiber tapers is also shown as the data point with the shortest wavelength. The inset shows the geometry simulated by FDTD, as well as the $|E|^2$ field of the cavity mode in the presence of the fiber taper.}
\label{fig:cavtune}
\end{figure}

Finally, we attempt to tune the wavelength of the nanobeam cavities by bringing a fiber taper in close proximity to the cavity. Fiber taper fabrication details and experimental setup are the same as in Ref. \cite{Gary_fiber}. In this configuration, we are able to both pump the cavity (at 780 nm) and collect the resulting 1.3 $\mu$m emission through the fiber. We observe the cavity spectrum as we move the fiber from a weak coupling position, where the cavity mode and fiber slightly overlap, to a strong coupling position, where the presence of the fiber red-shifts the cavity mode (as it has a higher index than air). In particular, we move the fiber along the $y$ and $z$ directions of Fig. \ref{fig:modes}(b), and plot the resulting spectra in Fig. \ref{fig:cavtune}(a). With the movements in the two directions, we demonstrate tuning of the cavity mode by over 7 nm. In addition, the $Q$ factor of the cavity is not greatly deteriorated throughout the movement of the fiber, as the free space measured $Q$ is $5,500$, while the measured $Q$ in the spectra in Fig. \ref{fig:cavtune}(a) range from 3,800 to 5,400. We simulate the full structure of the cavity with the fiber represented by a 1 $\mu$m diameter silica ($n=1.5$) cylinder lying horizontally along the nanobeam long axis (inset, Fig. \ref{fig:cavtune}(b)). We find via simulation that the fiber in full contact with the nanobeam redshifts the cavity resonance by 8.4 nm from the free space resonance wavelength, while supporting the same cavity mode with $Q$ = 15,000 ($Q$=35,000 without the fiber). These theoretical results match the experimental result of minimal perturbation of the cavity $Q$ in experiment, while the reduced tuning range of 7 nm could be due to imperfect positioning of the fiber on top of the cavity. If we were to use thinner membranes, the tuning range of the cavity mode could be even larger, as the design of the nanobeam cavity with $d/a=0.7$ confines the mode well in the $z$ direction. 

We also pump another cavity through the fiber taper and through free space (from normal incidence) with the 980 nm laser, and with the fiber taper in various positions near the cavity (in both cases, emission is collected via the fiber taper). We observe lasing in both cases, with thresholds obtained from a linear fit to the above threshold behavior, as before (Fig. \ref{fig:cavtune}(b)). The data point with the shortest wavelength represents the experiment without any fiber tapers, and we again observe that the cavity is red-shifted with the presence of the fiber taper. We also note that the fiber taper efficiently pumps the cavity, as we observe lower thresholds with the fiber taper pumping than with free space pumping. This efficient pumping is due to the localized pumping of the cavity from the fiber taper \cite{Lee_fibertaperlaser}. Furthermore, the free space pump case with the presence of the fiber taper has a much higher threshold than any other case, most likely due to the fiber taper reflecting or redirecting the pump away from the cavity. Finally, we notice that the threshold changes by more than a factor of two over a small wavelength range as the cavity wavelength is tuned by the movement of the fiber taper. Such a change is due to the change in pump efficiency, instead of the change in material absorption with wavelength. The taper is moved farther away from the center of the cavity as the cavity wavelength blue shifts, leading to reduced pumping of the cavity region and a higher threshold.

%tuned data from 2010/01/28, plotstuff.m
%more data from 2010/02/17

In conclusion, we have demonstrated CW lasing from InAs QDs embedded in 1D nanobeam PC cavities, with low thresholds of 0.3 $\mu$W and 19 $\mu$W for 780 nm and 980 nm pumping, respectively. Such a low threshold laser can be used in a variety of applications, such as optical communications or chemical sensing. We have also demonstrated that pumping below the GaAs band gap reduces heating of the cavity, while still enabling lasing.  We have investigated the wavelength dependence of the lasing threshold, with the increasing threshold for decreasing emission wavelength (approaching the peak of the QD distribution), resulting from higher material absorption. Finally, we have demonstrated that the nanobeam cavity laser can be tuned by over 7 nm by a fiber taper in close proximity to the cavity. The fiber taper can also serve as a pathway to efficiently pump the cavity.

The authors would like to acknowledge the MARCO Interconnect Focus Center, the NSF graduate research fellowship (YG, GS), and the Stanford Graduate Fellowship (BE) for funding. Fabrication was done at Stanford Nanofabrication Facilities.

\end{document}